\documentclass[conference,a4paper]{APSIPA2025}
\usepackage{amsmath}
\usepackage{graphicx}
\usepackage{multirow}
\usepackage{threeparttable}
\usepackage{comment}
\usepackage{bm}

\usepackage{bm,url,mathtools}
\usepackage{amsfonts}
\usepackage{booktabs}

\usepackage[subrefformat=parens]{subcaption}

\usepackage{geometry}
\usepackage{fancyhdr}

\fancypagestyle{firststyle}{
  \fancyhf{}
  \fancyhead[C]{2025 Asia Pacific Signal and Information Processing Association Annual Summit and Conference (APSIPA ASC)}
}

\newcommand{\apsipaspkdisthq}{
    \begin{figure}[t]
      \centering
        \includegraphics[width=1.\linewidth]{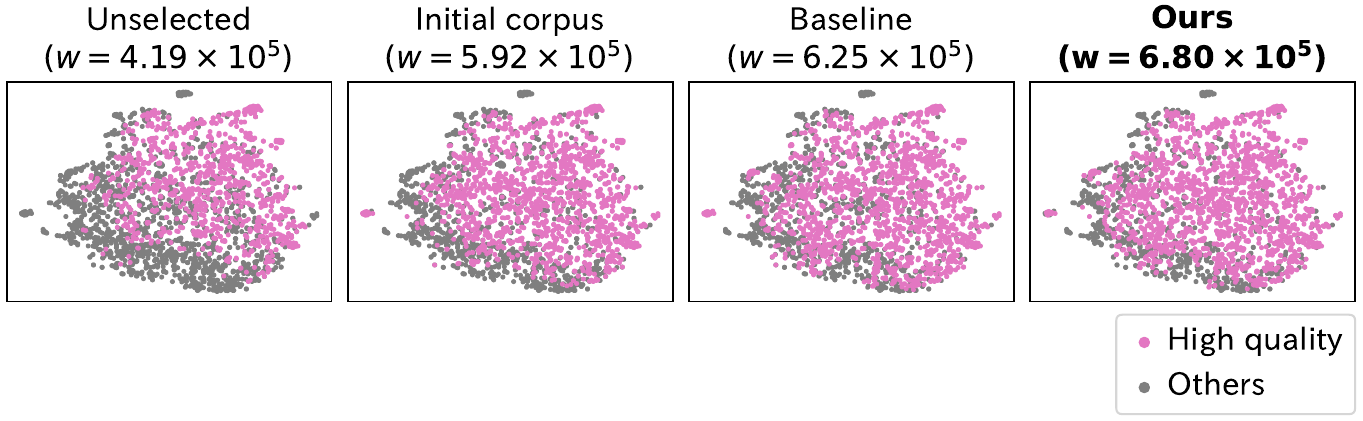}
      \vspace{-1mm}
     \caption{ 
        Distributions of high quality speakers by each data selection method.
        Higher $w$ indicates more diversity.
        Bold indicates the highest $w$ among the methods.
     }
    \label{fig:6:apsipahq_spk_dist}
     \vspace{-3mm}
    \end{figure}
}

\newcommand{\apsipasinglepseudomos}[1]{
    \begin{figure}[t]
    \centering
    \includegraphics[width=0.8\linewidth]{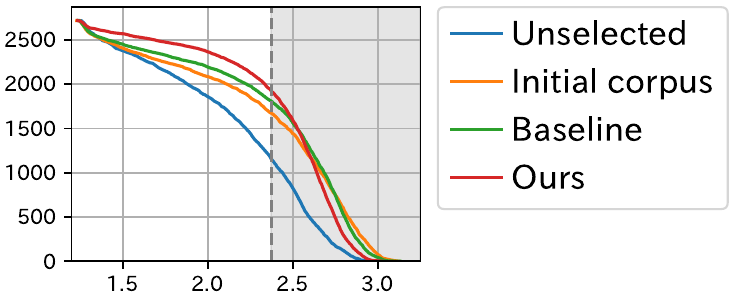}
    \caption{
        Cumulative histograms of pseudo MOS. 
        Y-axis value indicates number of speakers with higher score than x-axis value.
        The shaded area corresponds to high-quality speakers.
    }
    \label{fig:6:single_pseudo_mos}
    \end{figure}    
}

\newcommand{\apsipamatometepseudomos}[1]{
    \begin{figure}[t]
      \centering
      \begin{minipage}{\linewidth}
        \centering
        \includegraphics[width=0.8\linewidth]{chapter6/apsipa/single_pseudo_mos.pdf}
        \subcaption{Evaluation on real speakers.}
        \label{result_pseudmos_real}
      \end{minipage}
      \begin{minipage}{\linewidth}
        \centering
        \includegraphics[width=0.8\linewidth]{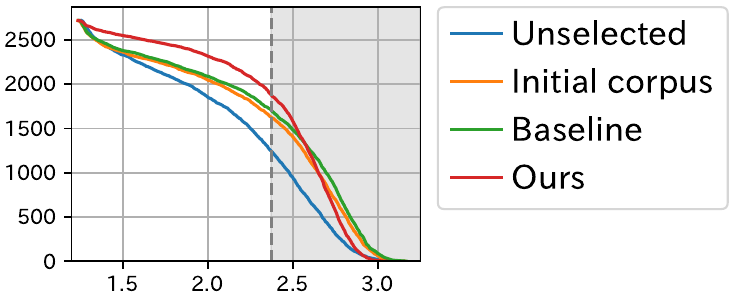}
        \subcaption{Evaluation on generated speakers.}
        \label{result_pseudmos_gen}
      \end{minipage}
    \caption{
        Cumulative histograms of pseudo MOS. 
        Y-axis value indicates number of speakers with higher score than x-axis value.
        The shaded area corresponds to high-quality speakers.
    }
    \label{fig:6:single_pseudo_mos}
    \end{figure}
}

\begin{document}

\title{Active Learning for Text-to-Speech Synthesis\\with Informative Sample Collection}

\author{
\authorblockN{
Kentaro Seki\authorrefmark{1}\authorrefmark{2},
Shinnosuke Takamichi\authorrefmark{1}\authorrefmark{2},
Takaaki Saeki\authorrefmark{1} and
Hiroshi Saruwatari\authorrefmark{1}
}
\authorblockA{
\authorrefmark{1}
The University of Tokyo, Japan.
\authorrefmark{2}
Keio University, Japan.\\
Email: seki-kentaro922@g.ecc.u-tokyo.ac.jp}

}

\maketitle
\thispagestyle{firststyle}
\pagestyle{fancy}

\begin{abstract}
The construction of high-quality datasets is a cornerstone of modern text-to-speech (TTS) systems.
However, the increasing scale of available data poses significant challenges, including storage constraints.
To address these issues, we propose a TTS corpus construction method based on active learning.
Unlike traditional feed-forward and model-agnostic corpus construction approaches, our method iteratively alternates between data collection and model training, thereby focusing on acquiring data that is more informative for model improvement.
This approach enables the construction of a data-efficient corpus.
Experimental results demonstrate that the corpus constructed using our method enables higher-quality speech synthesis than corpora of the same size.
\end{abstract}

\section{Introduction}
Recent advances in text-to-speech (TTS) synthesis have been strongly driven by the development of machine learning techniques that can leverage large-scale datasets~\cite{Tacotron2, MQTTS, ju2024naturalspeech}, making the construction of such datasets increasingly important. 
As the data requirements for TTS systems continue to grow, the mainstream approach to dataset construction is shifting from costly studio recordings~\cite{JSUT, JVS, take2024saslaw} to more scalable collection-based methods~\cite{LibriTTS, LibriVox_doitsugo, seki2023}. 
These methods typically involve collecting data from large-scale resources such as automatic speech recognition (ASR) corpus, audiobooks, and web data, and processing it into a format suitable for TTS model training.
Owing to the high degree of automation in data acquisition and preprocessing, they offer a significant advantage in terms of scalability.

Existing collection-based speech corpus construction approaches~\cite{chen2021gigaspeech, JTubeSpeech, nakata2024j} have primarily focused on collecting large quantities of speech data.
However, in practice, storage capacity constraint inevitably limit the amount of data that can be effectively utilized.
This makes it essential to consider how to efficiently select data that maximizes learning effectiveness within limited dataset size.
Nevertheless, many large-scale data collection strategies have paid insufficient attention to the optimization of data efficiency, leaving a critical aspect of practical TTS training unaddressed.

In connection with this issue, core-set selection methods have been proposed for TTS tasks~\cite{seki2024}.
Core-set selection aims to extract a representative subset from a large corpus that ideally achieves an equivalent learning effect as the original entire dataset~\cite{guo2022deepcore}.
Prior work~\cite{seki2024} improves data efficiency by maximizing diversity in the feature space and eliminating redundant samples; however, these methods are model-agnostic and do not take the learning dynamics of the TTS model into account.
In the field of image recognition, a previous study~\cite{ActiveLearningIsAStrongBaseline} have demonstrated that one of the strongest baselines for core-set selection is active learning, which iteratively refines both model training and dataset construction~\cite{ren2021survey}.
This suggests that incorporating active learning into core-set selection may enable the construction of more data-efficient TTS corpus.

In this study, we propose a data-efficient corpus construction method for TTS based on active learning.
Specifically, we focus on web-scale speech data, which is both large in volume and rich in diversity, and introduce a framework that selectively and incrementally utilizes such data.
Our method begins by listing and partitioning candidate web data sources, then sequentially collects each segment and trains a TTS model on it.
Based on the performance evaluation of the trained model, our method determines which parts of the next segment should be incorporated.
Furthermore, since the proposed approach downloads data segments on demand, it eliminates the need to download the entire dataset in advance, improving storage efficiency.
This feedback loop enables the model to focus on informative samples, thereby facilitating efficient and targeted corpus construction.
Experimental results show that the proposed method achieves superior speaker coverage compared to baseline methods, when trained on corpora of the same size.

\section{Proposed Method}

In this study, we focus on multi-speaker TTS models that are conditioned on $x$-vectors~\cite{xvector}, with the goal of enabling speech generation for a more diverse set of speakers.
Specifically, we define a speaker as being ``synthesizable'' if their synthetic speech exceeds a predefined quality threshold, and our method aims to increase the number of such synthesizable speakers.

\subsection{Preparation}
\subsubsection{Creating a Video ID list}
First, we create a list of video IDs $D_{all}$ to sample from YouTube.
This allows for fixing the distribution in subsequent random sampling, avoiding errors introduced by changes in distribution.
Video IDs are considerably lightweight compared to text-audio pairs, making it feasible to collect IDs for a larger number of videos than the text-audio pair collection.
In this study, we employ the method described in the previous study~\cite{seki2023} to search for videos and create a list of video IDs.
We then randomly shuffle the video ID list and divide it into $K$ disjoint subsets $D_1, D_2, ..., D_K$, each of which is used in a different iteration of data collection.  
The proportion of data used in the $k$-th iteration is denoted as $r_k := |D_k| / |D_{\mathrm{all}}|$, which is a predefined parameter that determines the sampling ratio for each iteration.

\subsubsection{Setting the target synthesis quality}
As stated at the beginning of this section, our goal is to increase the number of speakers for whom speech can be successfully synthesized.  
To this end, we define a quality threshold that serves as the criterion for determining whether a speaker is considered ``synthesizable.''
Specifically, we train a TTS model using an existing studio-recorded multi-speaker corpus and evaluate the synthesis quality for each speaker.  
We then define the minimum observed quality score among these speakers as the threshold $\theta_{hq}$.  
Speakers whose synthetic speech exceeds $\theta_{hq}$ are regarded as achieving a synthesis quality comparable to that of the studio-recorded corpus, and thus are considered capable of generating high-quality speech.

\subsection{Initial corpus construction}
\begin{figure}[t]
  \centering
    \includegraphics[width=0.8\linewidth]{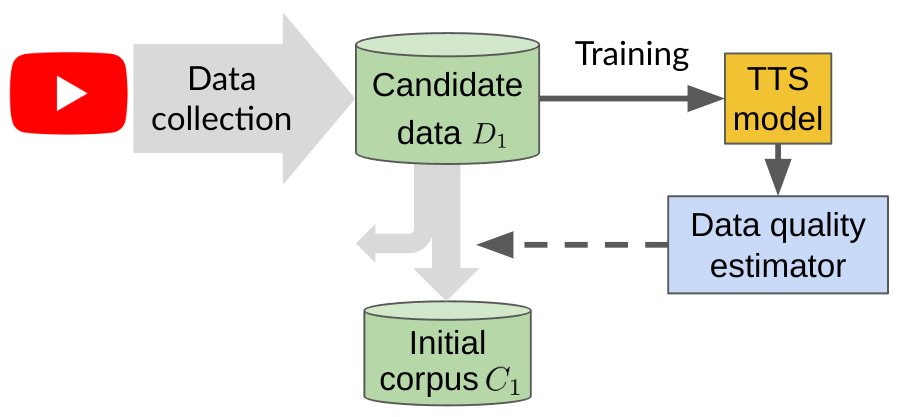}
  \caption{
    Overall procedure of initial corpus construction.
  }
  \label{fig:6:initial_corpus}
\end{figure}

\begin{figure}[t]
  \begin{minipage}{\linewidth}
    \centering
    \includegraphics[width=0.8\linewidth]{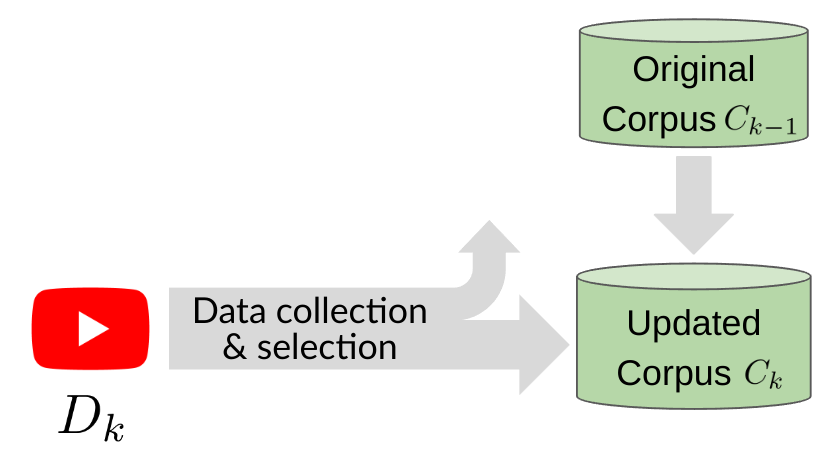}
    \subcaption{Additional data acquisition}
    \vspace{10pt}
    \label{fig:6:single-step_overall_construction}
  \end{minipage}
  \begin{minipage}{\linewidth}
    \centering
    \includegraphics[width=0.99\linewidth]{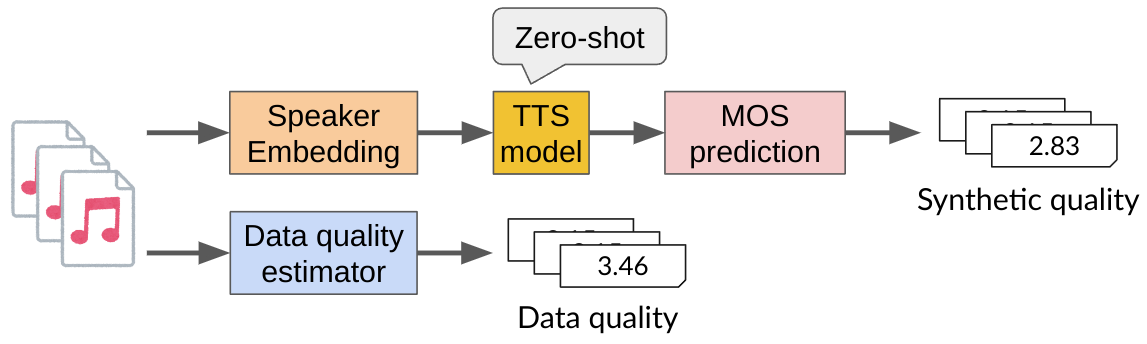}
    \subcaption{Data selection procedure}
    \label{fig:6:single-step_data-evaluation}
  \end{minipage}
  \caption{
    Overall procedure of additional data acquisition.
  }
  \label{fig:6:single-step}
\end{figure}

Fig.~\ref{fig:6:initial_corpus} shows the overall procedure of the initial corpus construction.

Following the previous study~\cite{seki2023}, we download audio-text pairs for the manually subtitled videos in $L_1$ and perform pre-screening based on two criteria: 
text-audio alignment accuracy and speaker compactness. 
Specifically, we compute the connectionist temporal classification (CTC) score~\cite{ctcsegmentation} to evaluate how well the audio aligns with the corresponding text, and filter out utterances with low alignment scores. 
Additionally, we estimate the intra-video variance of $x$-vectors~\cite{xvector} to assess speaker consistency within each video, discarding groups with high variance.
During the download process, we extract an $x$-vector for each video.
These $x$-vectors are lightweight compared to audio or text data and are saved for later use during inference.

We further need to filter out low-quality data from the downloaded set. To this end, we employ the training-evaluation loop method introduced in our previous study~\cite{seki2023, seki2025ttsops}. This method evaluates the quality of candidate training data by training a TTS model and then applying an automatic speech quality prediction model to the synthetic outputs. The evaluation model is trained to predict the synthesis quality of a TTS model that would be trained using the given data. Based on these predicted scores, we select utterances that are expected to exceed the threshold $\theta_{hq}$, thereby constructing the initial corpus $C_1$.

\subsection{Additional data acquisition}
As illustrated in Fig.~\ref{fig:6:single-step_overall_construction}, at the $k$-th step, we construct the corpus $C_k$ by selecting necessary data from $D_k$ and adding it to the existing corpus $C_{k-1}$.

First, we download audio-text data from $D_k$ in the same manner as described in Section~2.B.
Each data sample is then individually evaluated to determine whether it should be added to the corpus, based on two criteria: data quality and informativeness, as shown in Fig.~2b.

For data quality, we apply the data quality estimator trained in Section~2.B and select samples that exceed the threshold $\theta_{hq}$.

To assess informativeness, we train TTS model with the original corpus $C_{k-1}$.
Then, we extract the $x$-vector from each audio sample and perform zero-shot synthesis using this TTS model.
If the resulting synthetic speech exceeds the quality threshold $\theta_{hq}$, the sample is deemed redundant and excluded, as high-quality synthesis is already achievable without it.

In summary, we construct $C_k$ by adding to $C_{k-1}$ the samples from $D_k$ that satisfy both of the following: (1) data quality exceeds $\theta_{hq}$, and (2) synthetic quality falls below $\theta_{hq}$.

\subsection{Inference process}
The TTS system used in this study requires an $x$-vector as part of its input.
During the corpus construction, we performed a preprocessing step in which we extracted $x$-vectors from all videos in $D_{\text{all}}$.
As a result, we obtained and stored a set of speaker embeddings $\bm{x}_1, \dots, \bm{x}N \in \mathbb{R}^d$, where $d$ is the dimensionality of each $x$-vector and $N$ is the number of unique speakers in $D{\text{all}}$.
In the following, we describe how these precomputed $x$-vectors are used to infer the appropriate speaker identity at test time.

\subsubsection{Real speaker}
A straightforward strategy is to sample from the set $\left\{ \bm{x}i \right\}{i=1}^N$.
Since the corpus construction process aims to enable high-quality synthesis (above $\theta_{hq}$) for as many speakers in $D_{\text{all}}$ as possible, we expect that using these $x$-vectors will result in synthetic speech that meets or exceeds the threshold.

In cases where the synthesized speech for a speaker $\bm{x}i$ falls below $\theta_{hq}$, a plausible cause is that the corresponding training data were excluded due to its low quality.
While improving the quality of such data with data cleansing could enhance synthesis performance, this study does not address that issue.

\subsubsection{Generated speaker}
Since synthesis using real speaker $x$-vectors is inherently limited to the $N$ speakers observed in the dataset, we consider an alternative approach using generated $x$-vectors.  
This task is known as speaker generation (SG), and SG using Gaussian mixture models (GMMs) conditioned on speaker attributes has been previously proposed~\cite{tacospawn}.

In this study, the $x$-vectors are obtained by crawling speech data from the web and are not accompanied by explicit attribute annotations.  
Therefore, we adopt a data-driven approach to model the underlying structure of speaker embeddings using a diffusion model~\cite{ho2020denoising}. 
Diffusion models have been shown to effectively capture complex data distributions both theoretically~\cite{diffusion_models_are} and empirically~\cite{croitoru2023diffusion}, making them a suitable choice for modeling the diverse and intricate distribution of YouTube speaker data.

However, given that the number of available $x$-vectors ($N = 2719$ in our experiment) is small relative to their dimensionality ($d = 512$), we draw inspiration from latent diffusion models.  
Specifically, we apply principal component analysis (PCA) to decompose the embedding space and model only the principal components using a diffusion model, while approximating the remaining dimensions with a Gaussian distribution.

Let $\bm{\mu} \in \mathbb{R}^d$ and $\bm{R} \in \mathbb{R}^{d \times d}$ denote the empirical mean and covariance matrix of the $x$-vectors:
\begin{align}
\bm \mu &= \frac{1}{N} \sum_{i=1}^N \bm x_i, &
\bm R &= \frac{1}{N} \sum_{i=1}^N (\bm x_i - \bm \mu)(\bm x_i - \bm \mu)^T.
\end{align}
Let $\lambda_1, \ldots, \lambda_d$ be the eigenvalues of $\bm{R}$ in descending order, and let $\bm{e}_1, \ldots, \bm{e}_d$ be the corresponding orthonormal eigenvectors.  
We define $\bm{E} = [\bm{e}_1, \ldots, \bm{e}_d]$ and $\bm{\Lambda} = \text{diag}(\lambda_1, \ldots, \lambda_d)$.  
Then, we perform the following coordinate transformation to obtain a decomposition into principal and residual components:
\begin{align}
\begin{bmatrix} \bm y \\ \bm z \end{bmatrix} =
\bm{\Lambda}^{-1/2} \bm{E}^T (\bm{x} - \bm{\mu}),
\end{align}
where $\bm{y} \in \mathbb{R}^{d'}$ and $\bm{z} \in \mathbb{R}^{d - d'}$ for a chosen integer $d'$.  
We model $\bm{y}$ using a diffusion model and approximate $\bm{z}$ using a Gaussian distribution.  
By construction, $\bm{z}$ has zero mean and unit covariance, and is therefore modeled as $\bm{z} \sim \mathcal{N}(\bm{0}, \bm{I})$.
\section{Experimental evaluation}
\subsection{Experimental conditions}
\subsubsection{Data}
The data download and pre-screening procedures followed the same methodology as the prior study~\cite{seki2023}.
As a result, $2,719$ YouTube videos were processed, yielding approximately $66$ hours of candidate speech data, comprising around $60,000$ Japanese utterances.
Note that downloading all of them simultaneously was not required.
To calculate pseudo MOS, we used 100 phonetically balanced sentences from the JVS corpus~\cite{JVS}.
For the final evaluation of the trained TTS models, we used 324 test sentences from the ITA corpus~\cite{ita_git}.

\subsubsection{Text-to-speech model}
We employed FastSpeech 2~\cite{FastSpeech2} as our multi-speaker TTS model and used the UNIVERSAL\_V1 configuration of the pre-trained HiFi-GAN vocoder~\cite{Hifi-gan, hifigan_git}.
The model architecture and hyperparameters were adopted from the open-source implementation~\cite{FastSpeech2-JSUT}, with the exception of the speaker representation.
Instead of the original one-hot encoding for speaker identity, we utilized a publicly available $x$-vector extractor~\cite{jtube_xvector}. 
A $512$-dimensional $x$-vector was used to condition the TTS model, added to the encoder output via a $512$-by-$256$ linear transformation.
Each $x$-vector was computed by averaging over all utterances from the same speaker, resulting in one representative embedding per speaker.

\subsubsection{Speaker generation model}
We adopted $d'=28$ as the dimension of the diffusion model, determined by PCA to exceed a cumulative contribution rate of $99\%$. 
We concatenated the $28$-dimensional input with a $16$-dimensional embedding of $t$ and passed it through linear layer, ReLU, linear layer, ReLU, and linear layer to output $28$ dimensions. 
The hidden layer dimension was set to $56$.
We set time step $T$ to be $200$ and linearly increased $\beta_t$ from $\beta_1=0.0001$ to $\beta_T=0.05$. 
We split $x$-vectors from the $2719$ speakers into train, validation, and test sets in a ratio of approximately 8:1:1, resulting in $2175$, $272$, and $272$ speakers, respectively.

\subsubsection{Compared methods} 
We compared the following data selection methods.
\begin{itemize} \leftskip -5mm \itemsep -0mm
    \item \textbf{Unselected}: 
    All the collected data was used for the TTS training;
      the training data size was approximately $60,000$ utterances.
    \item \textbf{Initial corpus}:
    To examine the improvement achieved by data addition, we investigate the initial corpus $C_1$.
    \item \textbf{Baseline}:
    We execute the evaluation-in-the loop data selection described in the previous study~\cite{seki2023} with $n$ as a parameter corresponding to the corpus size.
    Specifically, $n$ was set to $3,943$, resulting in the same size as that of our proposed method.
    \item \textbf{Ours}:
    We construct the corpus using the proposed method.
    For the hyperparameter settings, 
      we set $K=2$, $r_1=0.1$ and $r_2=0.9$.
\end{itemize}

\subsubsection{Evaluation} 
\textbf{High-quality speakers (real speakers):}
We examine the number and distribution of speakers whose pseudo-MOS exceeded the threshold $\theta_{hq}$ and investigate whether the proposed method increases the number and distribution of high-quality speakers.
We conducted this investigation for both the speaker list and the generated speaker dataset.
The number of generated speakers is same as the speaker list, which is $2719$.

\textbf{High-quality speakers (generated speakers):}
We validate the effectiveness of the speaker generation model. 
First, we demonstrate that the diffusion model captures complex structures that cannot be captured by a GMM in the $x$-vector space. 
To achieve this, we generate a speaker dataset of the same size as the test subset and calculate the Wasserstein $1$-distance between the datasets. 
Since this value is a random variable that can vary depending on the generated dataset, we perform $30$ samplings to calculate the mean and standard deviation.
We adopt $M=1,2,...,10$ as the number of clusters for a GMM, and perform fitting for each.

\textbf{Generated speaker distribtion:}
To verify whether the quality evaluation of generated speakers is valid as an assessment of the synthesis quality of unseen speakers, we calculate the Wasserstein $1$-distance $d_{RG}$ between real speakers and generated speakers in the latent variable space, serving as an index of deviation from real speakers.
Since speaker characteristics are not always constant even for the same speaker, it is necessary to establish a reference value indicating that the generated speakers are different from real speakers. 
As a simple approach, one might consider calculating the Wasserstein-$1$distance $d_{RR}$ between different real speakers in the $x$-vector space. 
However, the preprocessing does not distinguish whether speakers from different videos are the same, making $d_{RR}$ inappropriate as a reference since it may inadvertently calculate the feature distances of different videos as if they were different speakers.
Therefore, in this study, we calculate the Wasserstein $1$-distance $d_{GG}$ between different generated speakers as the reference. 
As each generated speaker is independently produced, we can expect them to be different speakers. 
If $d_{RG}$ is of a similar magnitude to $d_{GG}$, we can consider the generated speakers as unseen speakers.

\subsection{Results}
\subsubsection{High-quality speakers}

\apsipamatometepseudomos

    \begin{table}[t]
    \caption{
        Ratio of high-quality speakers.
        Bold indicates the highest ratio in each block.
        
    }
    \label{table:6:matomete}
    \footnotesize
    \centering 
        \begin{tabular}{c|c||c|c}
            $n$ & Method & Real & Generated \\ \hline \hline 
            \multirow{1}{*}{$58500$}
  &Unselected & $42.6\%$ & $45.8\%$ \\ \midrule
\multirow{1}{*}{$2454$}
  &Initial corpus & $61.5\%$ & $59.8\%$ \\ \midrule
\multirow{2}{*}{$3943$}
  &Baseline & $66.6\%$ & $62.8\%$ \\ 
  &Ours & $\mathbf{71.0\%}$ & $\mathbf{69.2\%}$ 
        \end{tabular}
    \end{table}

\apsipaspkdisthq

Fig.~\ref{result_pseudmos_real} shows cumulative histograms of the pseudo MOSs, and Table~\ref{table:6:matomete} presents the ratio of high-quality speakers.
``Ours'' demonstrated a higher number of high-quality speakers compared to ``Baseline`` of the same size.
In addition, Fig.~\ref{fig:6:apsipahq_spk_dist} illustrates the distribution of high-quality speakers, where the spread is quantitatively represented by the value $w$.
Here, $w$ corresponds to the total length of the minimum spanning tree constructed from the $x$-vectors of the high-quality speakers, providing a measure of how widely the speakers are distributed in the embedding space.
These results indicate that our proposed methods are data-efficient approaches for corpus construction.
However, ``Ours'' demonstrated worse performance than ``Baseline`` in the range where pseudo MOS is $2.6$ or higher. 
We can say that this is because our proposed method does not further enhance speakers whose pseudo-MOS exceeds $\theta_{hq}$.

\subsubsection{High-quality speakers from generated speakers}
Fig.~\ref{result_pseudmos_gen} shows cumulative histograms of the pseudo MOSs with generated speakers.
The results are similar to the results with speaker list, and it can be said that ``Ours'' achieved more number of speakers whose synthetic quality is higher than $\theta_{hq}$.
This suggests that the TTS model trained in this experiment generalizes well to speakers and has the ability to synthesize even for speakers not included in the training data.

\subsubsection{Investigation for speaker generation model}
\begin{figure}[t]
\centering
\includegraphics[width=0.75\linewidth]{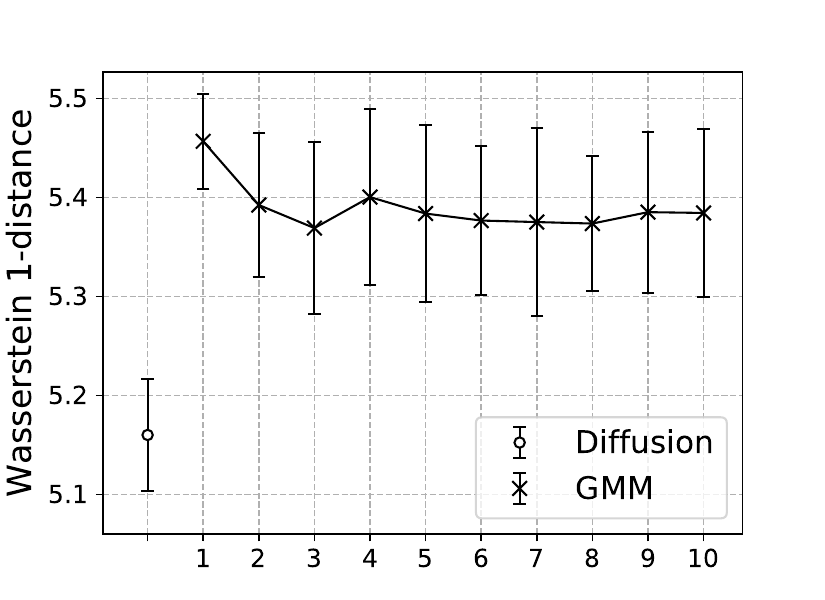}
 \caption{
    Wasserstein $1$-distance between the test dataset of speakers and generated speakers of the same size sampled from the speaker generation model. 
    The Circle represents the diffusion model, and crosses correspond to GMM. 
    The number of clusters for GMM is indicated on the x-axis. 
    Error bars represent twice the standard deviation.
  }
 \label{fig:6:wassaasutain}
\end{figure}

Fig.~\ref{fig:6:wassaasutain} shows the Wasserstein $1$-distance between test speakers and generated speakers sampled from the speaker generation models.
The error bars represent twice the standard deviation.
In all cluster numbers, the diffusion model is more than twice the standard deviation as distant from the GMM, indicating that the diffusion model captures the speaker distribution more appropriately.

$d_{RG}$, $d_{RR}$, and $d_{GG}$ took the following values, respectively.
\begin{align*}
 d_{RR} &=   3.492 \\
 d_{GG} &= 270.187 \\
 d_{RG} &= 268.382
\end{align*}
$d_{RR}$ is smaller than the other values, suggesting that the 2719 real speakers are actually counted as the same speaker with duplicates. 
Compared to $d_{RR}$, $d_{RG}$ takes values similar to $d_{GG}$. 
Therefore, we can conclude that the generated speakers are different from real speakers.

\begin{figure}[t]
\centering
\includegraphics[width=0.8\linewidth]{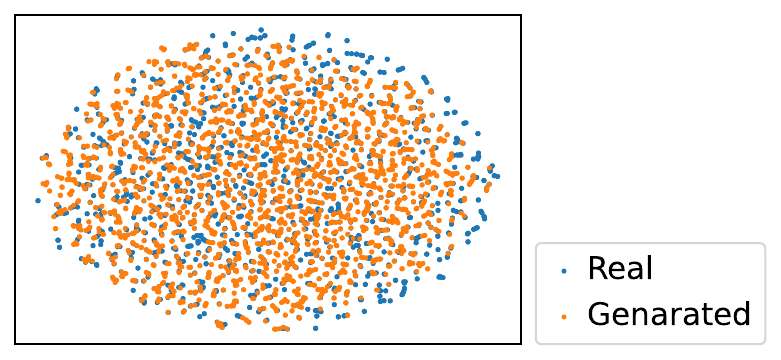}
 \caption{
    The results of dimensionality reduction using t-SNE for the $x$-vectors of real speakers and generated speakers.
    It can be observed that the x-vectors of generated speakers are distributed over a wide range.
  }
 \label{fig:6:latent_dist}
\end{figure}

To examine whether the generated speakers occupy the same embedding space as real speakers, we visualize the t-SNE projection of the combined $x$-vectors of both real and generated speakers, as shown in Fig. 6.
The visualization indicates that the generated speaker embeddings are distributed within the same region as those of real speakers.
Furthermore, the spread of the generated speaker embeddings demonstrates a diversity comparable to that of real speakers.
These observations suggest that the generated speakers are unseen yet lie within the same speaker space as real speakers, supporting the validity of the generation process.

\subsubsection{Investigation for data quality estimator}
\begin{figure}[t]
\centering
\includegraphics[width=0.6\linewidth]{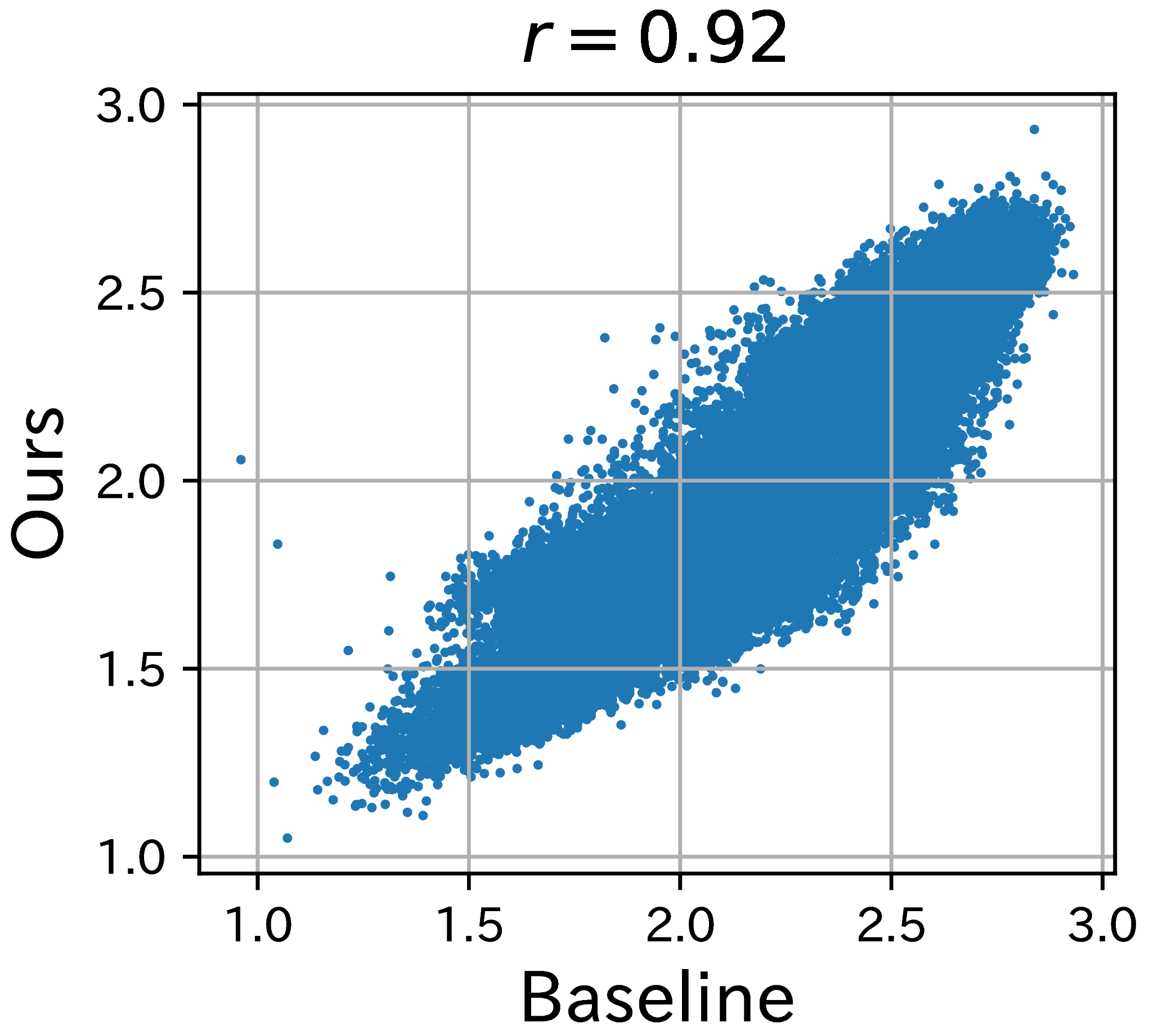}
 \caption{
    Comparison of evaluation model.
  }
 \label{evaluation_model_compare}
\end{figure}

Since our proposed method samples a subset of the candidate data for training the data quality estimator, it is important to investigate whether this modification affects the behavior of the estimator.  
Fig.~\ref{evaluation_model_compare} 7 shows a comparison between the estimators trained under two different conditions: the ``Baseline'' method, which uses all candidate data for training, and our proposed method (``Ours''), which uses only 10\% of that data.

The results indicate a strong correlation between the two estimators, suggesting that the proposed method maintains similar behavior despite the reduced training data. This finding implies that our approach effectively overcomes a critical limitation of the prior work—namely, the need to process the entire candidate dataset upfront to train the estimator.

\section{Conclusion}
In this study, we proposed a novel corpus construction framework for multi-speaker text-to-speech (TTS) synthesis based on active learning. Our method incrementally selects informative samples from large-scale web data, enabling data-efficient training without requiring the full dataset to be downloaded in advance. By integrating a data quality estimator and evaluating informativeness via zero-shot synthesis, the proposed method efficiently increases the number of high-quality synthesizable speakers.

Experimental results demonstrated that our method outperforms conventional baseline approaches in terms of speaker coverage and synthetic speech quality, even with the same corpus size. We also showed that the TTS model generalizes well to unseen speakers, and the diffusion-based speaker generation model captures the complex distribution of speaker embeddings more effectively than conventional GMMs. Additionally, we confirmed that the data quality estimator remains effective even when trained on only a fraction of the candidate data.

These findings suggest that our approach is a promising direction for scalable and practical corpus construction for TTS systems.

\section*{Acknowledgment}
This work was supported by JSPS KAKENHI 22H03639, 24KJ0860 and Moonshot R\&D Grant Number JPMJPS2011.











\bibliographystyle{IEEEtran}
\bibliography{references}

\end{document}